\documentclass[prl,twocolumn,tightenlines,preprintnumbers,floatfix,nofootinbib]{revtex4}
\usepackage{graphicx,amsfonts,amssymb,amsmath}

\def\be{\begin{equation}}
\def\ee{\end{equation}}
\def\ba{\begin{eqnarray}}
\def\ea{\end{eqnarray}}
\def\ge{\mathrel{\raise.3ex\hbox{$>$\kern-.75em\lower1ex\hbox{$\sim$}}}}
\def\la{\mathrel{\raise.3ex\hbox{$<$\kern-.75em\lower1ex\hbox{$\sim$}}}}

\def\simgt{\mathrel{\raise.3ex\hbox{$>$\kern-.75em\lower1ex\hbox{$\sim$}}}}
\def\simlt{\mathrel{\raise.3ex\hbox{$<$\kern-.75em\lower1ex\hbox{$\sim$}}}}

\newcommand{\bi}[1]{\bibitem{#1}}
\newcommand{\fr}[2]{\frac{#1}{#2}}

\newcommand{\nc}{\newcommand}

\nc{\gone}{\bar g_{\pi NN}^{(1)}}
\nc{\gzero}{\bar g_{\pi NN}^{(0)}}
\nc{\al}{\alpha}
\nc{\ga}{\gamma}
\nc{\de}{\delta}
\nc{\ep}{\epsilon}
\nc{\ze}{\zeta}
\nc{\et}{\eta}

\nc{\Th}{\Theta}
\nc{\ka}{\kappa}
\nc{\rh}{\rho}
\nc{\si}{\sigma}
\nc{\ta}{\tau}
\nc{\up}{\upsilon}
\nc{\ph}{\phi}
\nc{\ch}{\chi}
\nc{\ps}{\psi}
\nc{\om}{\omega}
\nc{\Ga}{\Gamma}
\nc{\De}{\Delta}
\nc{\La}{\Lambda}
\nc{\Si}{\Sigma}
\nc{\Up}{\Upsilon}
\nc{\Ph}{\Phi}
\nc{\Ps}{\Psi}
\nc{\Om}{\Omega}
\nc{\ptl}{\partial}
\nc{\del}{\nabla}
\nc{\ov}{\overline}
\nc{\newcaption}[1]{\centerline{\parbox{15cm}{\caption{#1}}}}
\nc{\hef}{$^4$He}
\nc{\het}{$^3$He}
\nc{\lisx}{$^6$Li}
\nc{\lisv}{$^7$Li}
\nc{\bes}{$^7$Be}
\nc{\beet}{$^8$Be}
\nc{\hefm}{^4{\rm He}}
\nc{\hetm}{^3{\rm He}}
\nc{\lisxm}{^6{\rm Li}}
\nc{\lisvm}{^7{\rm Li}}
\nc{\besm}{^7{\rm Be}}
\nc{\beetm}{^8{\rm Be}}
\nc{\bs}{(N$X^-$)}
\nc{\xm}{$X^-$}
\nc{\ben}{$^9$Be}
\nc{\benm}{^9{\rm Be}}

\def\beq{\begin{equation}}
\def\eeq{\end{equation}}
\def\bmat{\begin{displaymath}}
\def\emat{\end{displaymath}}
\def\bear{\begin{eqnarray}}
\def\eear{\end{eqnarray}}
\def\bery{\begin{array}}
\def\ery{\end{array}}
\def\bit{\begin{itemize}}
\def\eit{\end{itemize}}
\def\btab{\begin{tabular}}
\def\etab{\end{tabular}}
\def\btbl{\begin{table}}
\def\etbl{\end{table}}
\def\bfig{\begin{figure}[htb]}
\def\efig{\end{figure}}
\def\bpic{\begin{picture}}
\def\epic{\end{picture}}

\def\ga{\mathrel{\raise.3ex\hbox{$>$\kern-.75em\lower1ex\hbox{$\sim$}}}}
\def\la{\mathrel{\raise.3ex\hbox{$<$\kern-.75em\lower1ex\hbox{$\sim$}}}}
\def\gappeq{\mathrel{\rlap {\raise.5ex\hbox{$>$}}
{\lower.5ex\hbox{$\sim$}}}}
\def\lappeq{\mathrel{\rlap{\raise.5ex\hbox{$<$}}
{\lower.5ex\hbox{$\sim$}}}}

\def\gyr{{\rm \, G\kern-0.125em yr}}
\def\mev{{\rm \, Me\kern-0.125em V}}
\def\gev{{\rm \, Ge\kern-0.125em V}}
\def\tev{{\rm \, Te\kern-0.125em V}}

\begin{document}


\setcounter{page}{1}


\title{Bridging the primordial $A=8$ divide with Catalyzed Big Bang Nucleosynthesis}

\author{Maxim Pospelov$^{\,(a,b)}$}

\affiliation{$^{\,(a)}${\it Perimeter Institute for Theoretical Physics, Waterloo,
Ontario N2J 2W9, Canada}\\
$^{\,(b)}${\it Department of Physics and Astronomy, University of Victoria, 
     Victoria, BC, V8P 1A1 Canada}
}

\begin{abstract}

Catalysis of nuclear reactions by  metastable charged particles $X^-$
opens the possibility for primordial production of elements with $A>7$. 
We calculate the abundance of 
$\benm$, where synthesis is mediated by the formation of $(\beetm X^-)$ 
bound states, finding 
a dramatic enhancement over the standard BBN prediction: $\benm/^1{\rm H} \simeq 10^{-13}\times(Y_X/10^{-5})$.
Thus observations of \ben\ abundances 
at low metallicity offers a uniquely  sensitive probe 
of many particle physics models that predict $X^-$, 
including variants of supersymmetric models. Comparing the catalytically-enhanced 
abundances of primordial \lisx\ and \ben, 
we find the relation $\benm/\lisxm =$ $(2-5)\times 10^{-3}$
that holds over a wide range of $X^-$ abundances and lifetimes. 

\end{abstract}

\maketitle

\newpage

The first model of Big Bang Nucleosynthesis \cite{abg} (BBN) made an ambitious attempt
at explaining all elemental abundances 
as a result of successive primordial neutron capture. Two of the most important 
reasons why that theory did not work, and only very light nuclei 
can be generated in the Big Bang, are the nuclear $A=5$ and $A=8$ divides, 
or the absence of stable nuclear isotopes with these mass 
numbers. After five decades of progress in nuclear physics, astrophysics and cosmology, 
we now have a very successful framework of 
Standard Big Bang Nucleosynthesis (SBBN) that makes predictions 
for elemental abundances of light elements, H, D, He and Li, as functions of only one 
free parameter, the ratio of baryon to photon number densities \cite{Sarkar}.
With an additional 
CMB-derived \cite{WMAP} input value of $\eta_b=6\times 10^{-10}$, 
the comparison of observed amounts of deuterium, helium and lithium
serve as a stringent test of Big Bang cosmology and particle 
physics, and the importance of these tests 
is paramount for many theories seeking to extend the Standard Model
\cite{Sarkar}. 

Last year it was realized 
that BBN is sensitive to a wider class of particle 
physics models than was previously thought through the phenomenon of 
catalyzed primordial nuclear fusion \cite{Pospelov}, or CBBN. 
In particular it was shown that the
presence of metastable heavy negatively 
charged particles, that unavoidably will form bound states 
with nuclei \cite{Pospelov,BS,BS^2}, 
leads to the catalysis of \lisx\ production
via $ (\hefm X^-) +{\rm D}\to ~\lisxm +X^-$ 
\cite{Pospelov}, with a rate exceeding
the SBBN rate by many orders of magnitude. 
Moreover, the earlier presence of these particles at $T\sim 30$ keV 
leads to a moderate reduction of \lisv\ abundance \cite{Bird}, 
thus tantalizingly reproducing the observational pattern of 
\lisx\ and \lisv\ \cite{Litreview}.
Since then many groups have incorporated the catalyzed reactions
into their calculations \cite{others}, and the first 
dedicated three-body potential model calculation of the catalyzed 
rate was performed in \cite{3body}. 

The catalysis of \lisx\ 
can be viewed as an effective ``bridging" of the $A=5$ divide, 
and in this Letter we show that $A=8$ is also bridged
resulting in an enormous enhancement of the primordial \ben\ 
abundances over the SBBN value, $\benm/^1{\rm H}\la 10^{-18}$ \cite{Be_SBBN}, thus 
effectively incorporating \ben\ into the BBN ``family" of
light nuclei. As was pointed out in the original 
papers \cite{Pospelov,Bird}, the path to $A>8$ nuclei 
is controlled by the (\beet\xm) bound state, to get to which
\xm\ has to go through a ``double bottleneck" of
successive $\alpha$-captures: \xm$\to$(\hef\xm)$\to$(\beet\xm).
Once the (\beet\xm) bound state is formed, 
it participates in neutron capture, 
\begin{eqnarray}
&&\!\!\!\!\!\!\!\!\!\!\!\!\!\!\!\!\!
{\rm CBBN}:\,(\beetm X^-) +n\to\, \benm +X^-; ~Q_C = 0.26{\rm MeV},
\label{CBBN}
\end{eqnarray}
that is catalyzed relative to neutron capture on \beet, 
\begin{eqnarray}
&&\!\!\!\!\!\!\!\!\!\!\!\!\!\!\!\!{\rm SBBN}:\,  \beetm  +n\to\, \benm +\gamma; ~Q_S = 1.665{\rm MeV}.
\label{SBBN}
\end{eqnarray}
The latter is, of course, never important for BBN because free \beet\ lives for under a femtosecond. 
According to the general scaling of quantum mechanics \cite{Pospelov}, the cross section 
for (\ref{CBBN}) is enhanced relative to (\ref{SBBN}) by a large factor
$\sim c^3/(\omega a)^3$, where $\omega=Q_S/\hbar$ is the 
photon frequency in (\ref{SBBN}),
$a$ is the characteristic size of the (\beet\xm) system, and $c$ is the speed of light
($c=\hbar=1$ hereafter, and $\alpha=e^2$). In the rest of this Letter we analyze the rate for reaction (\ref{CBBN}) 
in detail, calculate the rates for (\beet\xm) formation, and incorporating them in 
the reaction network, calculate the resulting \lisx, \ben\ synthesis at $T\simeq 8$ keV
as a function of initial \xm abundance per baryon $Y_X$, and lifetime $\tau_X$. 
In this work we use the following input values for the r.m.s. charge radii $r^c_N$
of \hef, \beet, and \ben\, and calculate the binding energies of \bs\
in the limit of very heavy \xm  using a Gaussian charge distribution:
\begin{eqnarray}
r^c_{\hefm} = 1.67; & r^c_{\beetm}= 2.50 ;& r^c_{\benm} =2.50~{\rm ~~ fm}\\
E^b_{\hefm} = 347;  &E^b_{\beetm}=  1408;& E^b_{\benm} = 1477 ~{\rm  keV}.
\end{eqnarray}
While (\hef\xm) binding energy has an uncertainty under a few keV, 
the (correlated) uncertainty in $E^b_{\beetm}$ and $E^b_{\benm}$ can be as
large as 50 keV, if e.g. the charge distribution is varied from 
Gaussian to square. More precise values of binding 
energies are possible through theoretical and 
experimental advances in determining the electromagnetic form factors of light nuclei \cite{Sick}.

\begin{figure}[htbp]
\vspace{-1cm}
\centerline{\includegraphics[width=8cm]{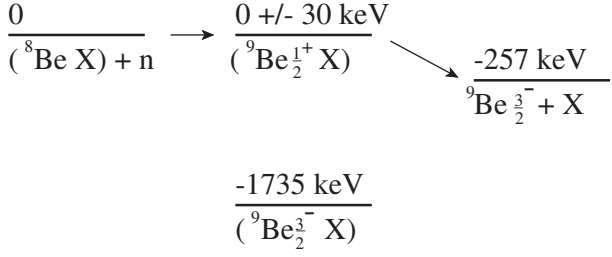}}
 \vspace{-0.7cm}\caption{
\footnotesize Isobar diagram for the $\beetm+n+X^-$ system.
The neutron capture occurs via a $(\benm\fr{1}{2}^+X^-)$ resonance
with subsequent decay to a $p$-wave continuum of \xm\ and  \ben\ in its ground state ($\fr32^-$)
releasing $Q = 257$ keV. The arrows show the path for the CBBN 
reaction that dominates over $\gamma$-emission to the ground state of
(\ben\xm) by four orders of magnitude. }
\label{Isobar_Be9} 
\end{figure}
{\em Catalyzed neutron capture}. It is well known that the $\beetm+n$ system has a 
low-energy neutron $s$-wave resonance ($E_n\simeq 70$keV  \cite{Be9res}) that 
corresponds to a $1/2^+$ excited state of \ben. 
Through the photodisintegration of \ben,
the energy and the width of this excited state 
are determined very accurately \cite{Be9res},
\be
E_{\fr{1}{2}^+}=1735\pm 3{\rm keV}; ~ \Gamma_n(E_n) \simeq 2\sqrt{\fr{192 E_n}{{\rm keV} }}; ~ 
\Gamma_\gamma= 0.57{\rm eV}.
\label{res_param}
\ee
Notice the $\sqrt{E}$ behavior of the width for the near-threshold resonance 
\cite{LL}.
In Fig.~\ref{Isobar_Be9} we show the isobar diagram for the $\beetm+n+X^-$ system,
that holds the key for determining the rate of reaction (\ref{CBBN}).
As is evident from this diagram, the neutron resonance occurs right at the threshold, 
$E_n^{\rm res} \sim 0 \pm 30$keV, with the estimated error related to differences in 
charge distribution for \ben\ and \beet. Fortunately, 
the neutron width in (\ref{res_param}) is larger than possible shifts of resonance energy, which 
makes this uncertainty tolerable for the calculation of 
the rate (even if the actual resonance is slightly below the threshold). 
Furthermore, there is no reason to assign a different width for the neutron resonance in
the $(\beetm X^-) + n$ system relative to $\beetm +n$ and we adopt $\Gamma_n(E_n)$ from
Eq. (\ref{res_param}). An accurate calculation of the decay 
width for the $(\benm\fr{1}{2}^+ X^-)\to \benm \fr32^- +X^-$ process requires a 
dedicated nuclear many-body calculation that we will not pursue in this paper. 
Instead, we choose to re-process the existing experimental information on $\Gamma_\gamma$ 
by extracting the strength of the reduced matrix element for the nuclear $E1$ transition,
\be
\label{Gg}
|\langle 1/2^+ \parallel d/e \parallel 3/2^-\rangle|^2 = 
\fr{3\Gamma_\gamma}{2\alpha \omega^3} = ({\rm 0.88~  fm})^2, 
\ee
and connect it with an approximate 
expression for the $(\benm\fr{1}{2}^+ X^-)$ $p$-wave decay width,
\begin{eqnarray}
\label{Gout}
\Gamma_{\rm out} &\simeq& \fr{\alpha^2}{2v}\times
|\langle 1/2^+ \parallel d/e \parallel 3/2^-\rangle|^2\times I_r^2;
\\\nonumber
I_r &=& \int r^2dr\times R_{10}(r)\fr{f(r)}{r^2} R_{p1}(r). 
\end{eqnarray}
In Eqs. ({\ref{Gg}) and (\ref{Gout}) $v$ is the velocity of the outgoing 
\ben\ in the catalyzed reaction, $I_r$ is the radial integral with dimension 
[distance]$^{-3/2}$, $R_{10}$ is the ``atomic" $1s$ radial wave function of 
the decaying (\ben\xm) system, $R_{p1}$ is the final state $l=1$ wave function with 
$p=\sqrt{2m_{\benm} Q}$, normalized to the $p/2\pi$ scale \cite{LL}, that in large $r$ limit becomes 
the wave function for the Coulomb problem. $f(r)$ represents a ``form factor" that 
we have to assign to the interaction of a nuclear dipole operator with the electric field created 
by \xm, such that $f(r)\rightarrow 1$ if the orbit of \ben\ were far from \xm.
For our calculation, we 
take $f(r) = 4\pi\int^r_0x^2dx \rho_c(x)/e$, so that for a constant 
charge density it scales as $f\sim r^3$ inside the 
nuclear radius and $f=1$ outside. Explicit numerical calculation gives
\be
\label{radial}
I_r \simeq(5.0~{\rm fm})^{-3/2}.
\ee
We note the resulting length scale in (\ref{radial}) is six times larger than 
the effective size of the nuclear dipole in (\ref{Gg}) giving some 
{\em a posteriori} justification to our procedure. Combining (\ref{Gg}), (\ref{Gout}),
and (\ref{radial}) we arrive at the following estimate for the decay width, 
\be
\Gamma_{\rm out} = \fr{3\alpha}{4v}~ \fr{\Gamma_\gamma I^2_r}{\omega^3} \simeq 5~{\rm keV},
\label{Gout_number}
\ee
which constitutes four orders of magnitude enhancement over $\Gamma_\gamma$. 
This estimate looks natural, perhaps on the lower side, 
given that $\Gamma_{\rm out}$ contains no small parameters
and the energy level spacing in this system is on the order of 1.5 MeV. 
Defining $\Gamma_{\rm tot}(E_n) = \Gamma_{\rm out} +\Gamma_n(E_n)$, we
calculate the cross section of catalyzed photonless neutron capture (\ref{CBBN})
using the Breit-Wigner formula, 
\be
\sigma_n(E_n) = \fr{g\pi}{k^2_n} \fr{\Gamma_n\Gamma_{\rm out}}{(E-E_n^{\rm res})^2 +\Gamma^2_{\rm tot}/4}.
\ee
$g$ is the spin factor, and  $g=1$ for (\ref{CBBN}).
The result is a factor of $\sim 5$ lower than the unitarity limit 
at $E\sim 10$keV. Thus at temperatures $T/10^9{\rm K} =T_9 = 0.1$ the reaction rate 
is calculated to be 
\be
 N_A\langle \sigma_nv_n\rangle \simeq 2 \times 10^9 ~ {\rm cm}^3{\rm mol}^{-1}{\rm s}^{-1},
 \label{estimate}
\ee
which is large but not larger  than {\em e.g.}
the rate of neutron capture on \bes. Since the actual 
position of $E_n^{\rm res}$ has an uncertainty of 
$\pm 30$keV we do not have sufficient precision to obtain the 
variation of the rate with temperature. We did check, however, that variation in the 
resonant energy by $\sim 30$ keV introduces only moderate shifts of a factor of $\sim 2$. 
Variations in $ \Gamma_{\rm out}$ directly affect the rate. 
A moderate increase of $\Gamma_{\rm out}$ 
by a factor of $4$ would increase the cross section 
to near the unitarity bound while
a decrease of $\Gamma_{\rm out}$ will obviously have an opposite effect.
Fortunately, dedicated {\em ab-initio}
calculations of $\Gamma_n(E_n)$ and 
$\Gamma_{\rm out}$ are feasible in 
state-of-the-art nuclear physics \cite{GFMC}, 
with the clear potential of improving the accuracy of estimate (\ref{estimate}).

\begin{table}
\begin{center}
\begin{tabular}{cccc}
\hline
~~~~$nl$~~~~ & ~$E_{nl}$(keV)~ &  ~$E_{\rm res}$(keV)~ & ~$\Gamma_\gamma$(eV)~ \\ \hline\hline
3$s$&          $-265$  &   173 &     0.1\\ 
3$p$&          $-323$ &   114 &   1.1 \\
3$d$&          $-351$ &   88 &      1.0  \\
2$s$&         $- 524$ &  $-86$ &     0.5 \\ 
2$p$&         $-706$  &  $-267$ &     4.5 \\ 
1$s$&        $-1408$  &    --  &    --   \\ \hline
\end{tabular}
\end{center}
\caption{ Properties of (\beet\xm) bound states with respect to $nl$ quantum numbers. 
Binding energies are 
shown relative to the \beet $+$\xm continuum. Resonant energies are given for the (\hef\xm)$+$\hef\ system. 
}
\vspace{-0.5cm}
\label{table1}
\end{table}
{\em Formation of {\rm(\beet\xm)}.} Before neutrons can undergo capture and form 
\ben, first the bound states (\hef\xm) and then (\beet\xm) 
have to form, so that the path to \ben\ is guarded by the double bottleneck. 
There are two main paths to
(\beet\xm): linear in $Y_X$ due to the $\alpha$-reaction on (\hef\xm); and quadratic 
in $Y_X$ due to formation of the neutral molecules (\hef\xm\xm) that react with helium 
via a strong Coulomb-unsuppressed process $(\hefm X^-X^-)+\hefm\to (\beetm X^-)+X^-$. 
Both channels first require the formation of 
(\hef\xm) that cannot form in any significant amounts above 10 keV due to photodissociation. 
We consider these mechanisms in turn. 

In the interesting range of temperatures, the radiative fusion of 
\hef\ on (\hef\xm) proceeds via the formation of the resonant atomic states
with $n=3$, similar to what is found in \cite{Bird}. 
\beet\ has an energy excess of 92keV relative to two alpha particles. This 
leaves only the $n=1,2$ levels below the threshold of (\hef\xm)+\hef\ continuum. 
The energy levels of the relevant (\ben\xm) excitations, resonant energies for incoming 
$\alpha$-particles and the electromagnetic decay widths are given in Table 1. 
The key observation that  facilitates the whole treatment is the validity of 
the narrow resonance approximation. By an appropriate  
rescaling of the Gamow factor for free \beet\ decay, it is easy to see 
that at relevant energies the entrance widths $\Gamma_{\rm in}$
to $3l$ states are $\sim O({\rm keV})$ and thus satisfy an important condition,
$\Gamma_\gamma \ll \Gamma_{\rm in} \ll T,$
which makes the cross section independent of $\Gamma_{\rm in}$ 
and totally determined by $E_{\rm res}$ and $\Gamma_\gamma$. Choosing $g=2l+1$ in 
the Breit-Wigner formula, and retaining the contributions from the two most important  resonances, 
$3d$ and $3p$, we 
derive the thermal rate $\langle \sigma v\rangle$ 
for the $(\hefm X^-)+\hefm\to (\beetm X^-)+\gamma$ reaction ($Q=969$keV) to be
\be
\label{res_rate}
10^5T_9^{-3/2}\left(0.95\exp[-1.02/T_9]
+0.66\exp[-1.32/T_9]\right).
\ee
The non-resonant part of the rate can be somewhat enhanced 
compared to a typical $(\alpha,\gamma)$ reaction
due to the subthreshold $2l$ resonances, but 
at $T\sim 10$keV it is totally negligible in comparison 
with (\ref{res_rate}). At these temperatures and with this rate the capture of \hef\ 
on (\hef \xm) is about two orders of magnitude slower than the Hubble rate
and rapidly dropping with $T$. This is also important, as it shows that only a 
relatively small but non-negligible fraction of (\hef\xm) will be converted to 
(\beet\xm).

The rate depends very sensitively on the $3p$ and $3d$ 
energy levels, but subtleties of the charge distribution in \beet\ make
little difference for their energies. 
We find it remarkable that one can calculate the abundance of (\beet\xm) virtually 
free of major nuclear physics uncertainties. 
In fact, the main correction to (\ref{res_rate}) comes from the $m_{\beetm}/m_{X}$-suppressed contribution to 
resonant energies that we ignore in this paper,  but can account for very easily.
For \xm as light as 100 GeV, these corrections amount to a 12 keV upward shift of
$E_{\rm res}$, resulting in a factor of a few suppression in (\beet\xm) abundances,
but quickly become negligible for heavier $m_X$.

The molecular mechanism of forming (\beet\xm) is completely different,
as it is regulated by  Coulomb unsuppressed processes. 
Due to the $Y_X^2$ scaling, this mechanism is 
 of secondary importance because $Y_X$ is rather tightly constrained by \lisx\ overproduction \cite{Pospelov}.
We calculate the rate of molecular formation $(\hefm X^-)+X^-\to (\hefm X^-X^-)+\gamma,~~Q\simeq 320$keV,
in the spirit of Kramers and ter Haar \cite{KtH}, treating
the ``nuclear" motion of \xm\ semiclassically, and ``electron" motion of 
\hef\ quantum-mechanically, which is known to give a good approximation 
to a full quantum mechanical treatment. At temperatures $T_9\ll Q$ one finds 
\be
\langle \sigma v \rangle_{\rm mol} =  8\pi^{1/2}T^{-1/2}\int r^2dr \times \Gamma_\gamma(r)|V(r)|^{1/2},
\ee
where $r$ is the distance between two \xm\ particles, 
$V(r)$ is the potential energy of (\hef\xm)--\xm\ interaction,
and $\Gamma_\gamma(r)$ is the probability per time for a quantum jump 
of \hef\ from the atomic to the molecular state with emission of a photon. 
Using the variationally determined molecular wave functions, we 
calculate $\Gamma_\gamma(r)$, $V(r)$ and find
the following estimate for the rate of molecular formation 
\be
\label{molecular}
N_A\langle \sigma v \rangle_{\rm mol} \sim 40\times T_9^{-1/2}~ {\rm cm}^3{\rm mol}^{-1}{\rm s}^{-1}.
\ee
Note that the molecular rate is significantly smaller than the ``atomic" (\hef\xm) recombination rate, 
$8\times 10^3T_9^{-1/2}$, but is not suppressed by $m_{\hefm}/m_X$, which is a direct consequence of 
the Coulomb attraction in the initial state.

{\em Synthesis of {\rm \ben}.} Rates (\ref{estimate}), (\ref{res_rate}), and (\ref{molecular})
enable us to calculate the \ben\ freezeout abundance numerically. 
Before we do that, we would like to mention that in the 
hypothetical limit of both 
$Y_X$ and $\tau_X$ being large the production of \ben\ 
will be neutron-supply-limited, and {\em all neutrons} produced in the 
DD and DT fusion below $T_9=0.1$ may end up captured by (\beet\xm) before they decay,
leading to  \ben[large $Y_X,\tau_X$]$\ga O(10^{-9})$. 
Given that at lowest metallicities a $10^{-14}-10^{-13}$ range for \ben\ 
is being probed \cite{Be9_obs}, this would constitute gross 
overproduction of \ben, which is clearly excluded. 

Figure \ref{Be9Li6} gives the result of our 
CBBN calculation at $T_9 < 0.12$. Besides rates 
(\ref{estimate}), (\ref{res_rate}), and (\ref{molecular}) calculated 
in this paper, we include the \lisx\ CBBN rate \cite{Pospelov} with the 
use of the $S$-factor properly calculated in \cite{3body}. 
\begin{figure}[htbp]
\vspace{-0.5cm}
\centerline{ \vspace{-0.2cm}\includegraphics[width=8.3cm]{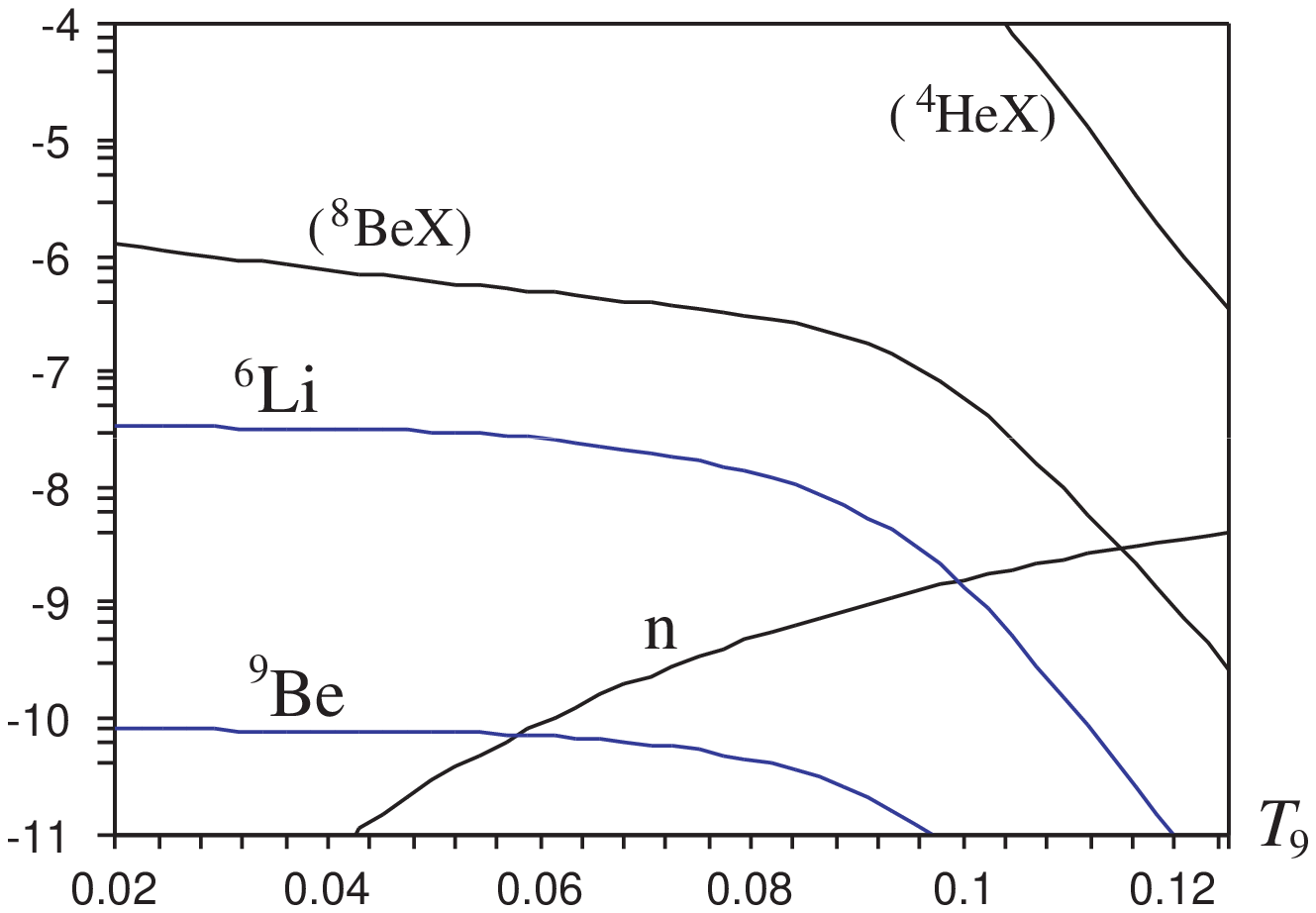}}
\vspace{-0.3cm}
\hspace{0.25cm}\centerline{\includegraphics[width=8.2cm]{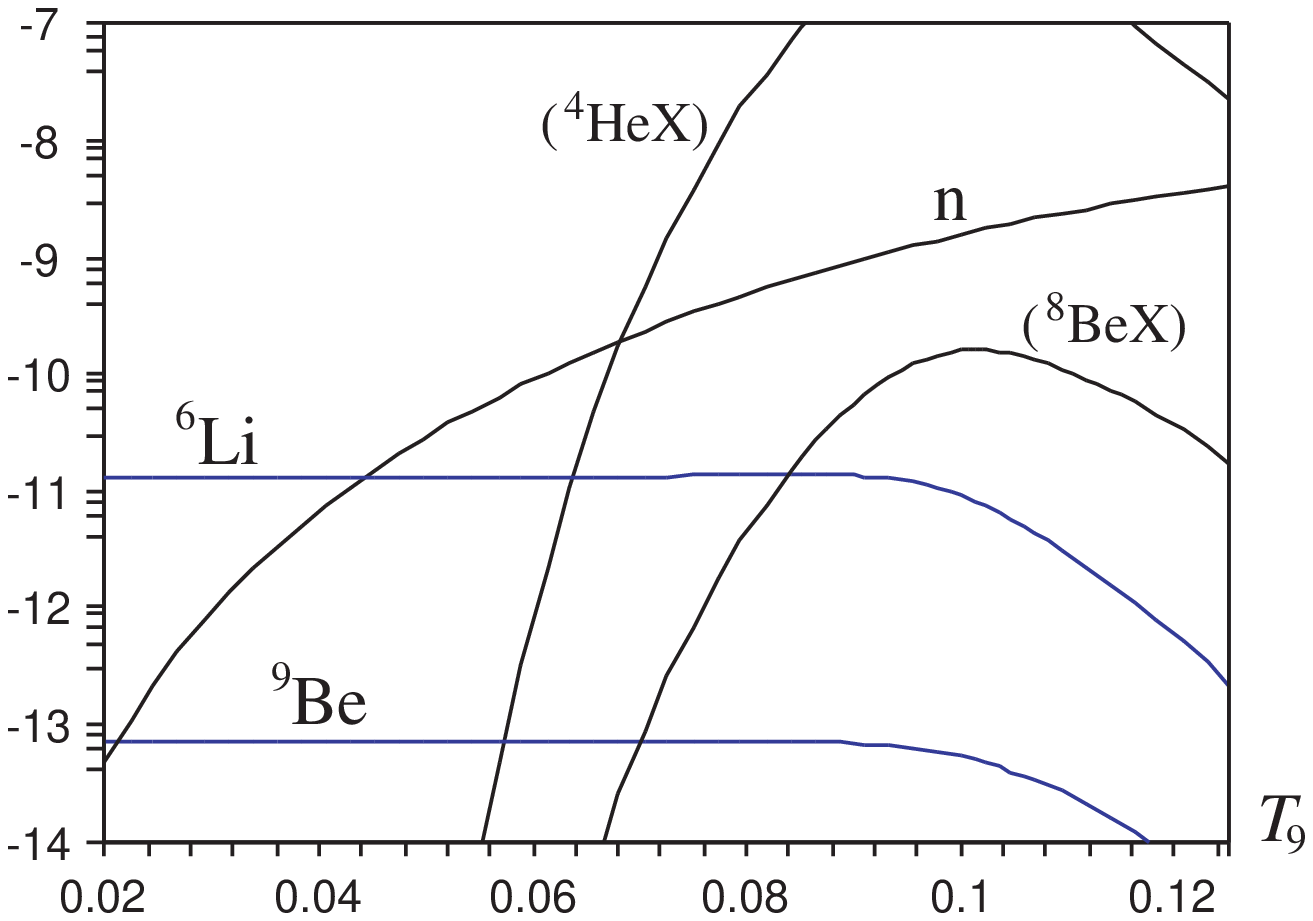}}
\caption{
\footnotesize Log$_{10}$ of the elemental abundances in CBBN for two choices of $\{Y_X,\tau_X\}$
input. The top figure represents very long lifetime $\tau_X$ and abundance of $Y_X=5\times 10^{-3}$. 
This option is excluded because \ben\ is produced with $O(10^{-11}-10^{-10})$ abundance. 
The increase in (\beet\xm) at low $T$ is due to continuing molecular formation. 
The lower plot represents $\tau_X=2000$s and initial abundance of $Y_X=0.1$,
suggested by the solution to the \lisv\ problem \cite{Bird}. For these parameters $\lisxm= 1.3\times 10^{-11}$, 
and $\benm=7\times 10^{-14}$. }
\label{Be9Li6} 
\end{figure}

The main result of this calculation is summarized as: 
\be
\benm \simeq 10^{-13} \times \left[Y_X(t=2\times 10^4{\rm sec})/10^{-5}\right].
\label{bound} 
\ee
Given that in some cases  \ben\ is detected below $\times 10^{-13}$ \cite{Be9_obs}, and typical
abundances $Y_X \ga O(10^{-3})$ are expected from annihilation of $X^-X^+$ at the freezeout, this 
 restricts the lifetime of \xm\ to a few thousand seconds, 
 reinforcing the lithium bound \cite{Pospelov}. It is also important
 that \ben\ is far less fragile than \lisx\ and therefore is unlikely to experience a 
 significant reduction of its abundance in subsequent evolution after the Big Bang,
making prediction (\ref{bound}) especially valuable.  
 Close inspection of Fig. \ref{Be9Li6} reveals a very similar behavior for 
 \lisx\ and \ben. Dividing the two abundances, we eliminate the dependence on 
 $Y_X$, obtaining the following relation, 
 \be
\benm_{\rm primordial} /\lisxm_{\rm primordial}  \simeq (2-5)\times 10^{-3},
 \ee
which is valid as long as $\tau_X \ga 2000$s. 
Intriguingly, this is exactly what is observed \cite{Be9_obs,Li6}, if we interpret the \lisx\ results as 
a ``primordial plateau", and take seriously [very tenuous] hints of elevated levels of \ben\ at lowest 
metallicities. 
It is of course possible that all abnormalities 
observed in lithium and beryllium abundances will 
find astrophysical explanations 
having nothing to do with the Big Bang \cite{keith_french},
and only more theoretical and observational work in this direction will clarify this issue. 

To conclude, metastable negatively charged particles 
are predicted by many particle physics models, 
including some variants of supersymmetry. These particles,
should they live in excess of 1000s, trigger CBBN and in this 
paper we show that sizable amounts of \ben\ can be generated,
and calculated with reasonable accuracy from first principles.
Somewhat reduced \lisv\ abundance \cite{Bird}, strongly enhanced 
primordial values of \lisx\ and \ben, and  \ben/\lisx $\sim$few$\times 10^{-3}$  
constitute a typical ``footprint" of CBBN. In this light, further 
observational studies of \ben\ and \lisx\ at low metallicities 
find an unexpected and very strong motivation from modern particle 
physics.

\end{document}